\documentclass[aps,prb,english,amsmath,amsfonts,amssymb,superscriptaddress,twocolumn,showpacs,citeautoscript,reprint,longbibliography]{revtex4-1}
\usepackage[T1]{fontenc}
\usepackage[latin9]{inputenc}
\usepackage{amsmath}
\usepackage{amssymb}
\usepackage{wasysym}
\usepackage{adjustbox}
\usepackage{esint}
\usepackage{graphicx}
\usepackage{times}
\usepackage{nicefrac}
\usepackage{appendix}
\usepackage{multirow}
\usepackage{textcase}
\usepackage{subfigure}
\usepackage{empheq}
\usepackage{color}
\usepackage{leftidx}
\usepackage{hhline}
\usepackage{makecell}
\usepackage{stackrel}
\usepackage{lineno}
\makeatletter
\@ifundefined{textcolor}{}
{%
 \definecolor{BLACK}{gray}{0}
 \definecolor{WHITE}{gray}{1}
 \definecolor{RED}{rgb}{1,0,0}
 \definecolor{GREEN}{rgb}{0,1,0}
 \definecolor{BLUE}{rgb}{0,0,1}
 \definecolor{CYAN}{cmyk}{1,0,0,0}
 \definecolor{MAGENTA}{cmyk}{0,1,0,0}
 \definecolor{YELLOW}{cmyk}{0,0,1,0}
}

\renewcommand{\Im}{\operatorname{Im}}

\newcommand{\asym}{\operatorname{asym}}

\renewcommand{\b}{\beta}

\newcommand{\add}[1]{\if\a\b{{\color{red} #1}}\else{#1}\fi}

\newcommand{\bracket}[1]{\langle #1 \rangle}
\newcommand{\ket}[1]{| #1 \rangle}
\newcommand{\bra}[1]{\langle #1 |}
\newcommand{\im}{\operatorname{i}}

\newcommand{\citeasnoun}[1]{Ref.~\onlinecite{#1}}

\renewcommand{\eqref}[1]{(\ref{eq:#1})}

\newcommand{\figref}[1]{Fig.~\ref{fig:#1}}

\newcommand{\appref}[1]{Appendix~\ref{sec:#1}}

\newcommand{\Secref}[1]{Section~\ref{sec:#1}}

\newcommand{\bs}[1]{\boldsymbol{#1}}
\newcommand{\trace}[1]{{\rm Tr} \left[ #1 \right]}

\makeatother
\usepackage{babel}

\begin{document}
\title{Channel-based algebraic limits to conductive heat transfer}

\author{Prashanth S. Venkataram}
\author{Sean Molesky}
\affiliation{Department of Electrical Engineering, Princeton
  University, Princeton, New Jersey 08544, USA}

\author{Juan Carlos Cuevas}
\affiliation{Departamento de F\'{\i}sica Te\'orica de la Materia
  Condensada and Condensed Matter Physics Center (IFIMAC), Universidad
  Aut\'onoma de Madrid, E-28049 Madrid, Spain}

\author{Alejandro W. Rodriguez}
\affiliation{Department of Electrical Engineering, Princeton
  University, Princeton, New Jersey 08544, USA}

\date{\today}

\begin{abstract}
  Recent experimental advances probing coherent phonon and electron
  transport in nanoscale devices at contact have motivated theoretical
  channel-based analyses of conduction based on the nonequilibrium
  Green's function formalism. The transmission through each channel
  has been known to be bounded above by unity, yet actual
  transmissions in typical systems often fall far below these
  limits. Building upon recently derived radiative heat transfer
  limits and a unified formalism characterizing heat transport for
  arbitrary bosonic systems in the linear regime, we propose new
  bounds on conductive heat transfer. In particular, we demonstrate
  that our limits are typically far tighter than the Landauer limits
  per channel and are close to actual transmission eigenvalues by
  examining a model of phonon conduction in a 1-dimensional chain. Our
  limits have ramifications for designing molecular junctions to
  optimize conduction.
\end{abstract}

\maketitle 


Tailoring nanoscale devices for conductive heat transfer (CHT) is
relevant to the design of thermoelectric devices, heat sinks and
refrigerators, optoelectronic and optomechanical devices, and for
control over chemical reactions at the nanoscale~\cite{SegalARPC2016,
  TianPRB2012, TianPRB2014, BurklePRB2015, KlocknerPRB2017A,
  KlocknerPRB2017B, KlocknerPRB2017C, LuoPCCP2013, CahillAPR2014,
  PopNANORES2010}. Recent experiments have accurately measured thermal
conductance through single-atom or molecular
junctions~\cite{CuiSCIENCE2017, MossoNATURE2017, CuiNATURE2019,
  MossoNANOLETT2019}. Concurrently, phonon and electron conduction
have been theoretically described in the linear response regime via
the nonequilibrium Green's function method~\cite{CuevasPRL1998,
  SegalARPC2016, MingoPRB2003, TianPRB2012, TianPRB2014, DharJSP2006,
  BurklePRB2015, KlocknerPRB2016, KlocknerPRB2017A, KlocknerPRB2017B,
  KlocknerPRB2017C, KlocknerPRB2018, ZhangPRB2018, SadasivamPRB2017,
  DubiRMP2009}. Many of these theoretical works have studied
conduction from the perspective of channel-based transmission
contributions at each frequency. The transmission from each channel is
theoretically bounded above by unity, but in typical systems, the
actual transmission falls far short of these limits, and it has been
difficult to produce general predictive or explanatory insights
regarding which systems may or may not exhibit transmission
contributions close to these bounds~\cite{KlocknerPRB2017B}.

In this paper, building upon an accompanying
manuscript~\cite{VenkataramARXIV2020}, we provide channel-based upper
bounds for heat transfer, including CHT, in arbitrary linear bosonic
systems that are at least as tight as Landauer limits at each
frequency~\cite{BurklePRB2015, KlocknerPRB2016, KlocknerPRB2017A,
  KlocknerPRB2017B, KlocknerPRB2017C, KlocknerPRB2018}, and in
practice are much tighter. First, in~\Secref{rankbounds}, we prove
that for the typical system corresponding to a junction connecting two
leads, the number of nonzero transmission eigenvalues at each
frequency is bounded above by the rank of the response of the junction
dressed by the two leads, and this in turn is bounded above by the
narrowest bottleneck in the junction. This derivation serves as an
algebraic proof or a prior statement based on physical albeit
heuristic arguments~\cite{CuevasPRL1998, BurklePRB2015,
  KlocknerPRB2016, KlocknerPRB2017A, KlocknerPRB2017B,
  KlocknerPRB2017C, KlocknerPRB2018}. Second,
in~\Secref{channelbounds}, we show that a recently derived unified
framework for heat transfer based on the non-equilibrium Green's
function formalism~\cite{VenkataramARXIV2020} can be used to
generalize recently proposed bounds on radiative heat transfer
(RHT)~\cite{MoleskyPRB2020, VenkataramPRL2020} to include phonon
transport; the derivation of these bounds is extremely technical and
complex, but we include it in the main text for completeness, and
point readers to~\eqref{Phiopt} for the main result. Third,
in~\Secref{results}, we demonstrate the much greater tightness of
these bounds for phonon CHT (i.e. coherent thermal phonon transport),
relative to the Landauer limits of unity, in a model 1-dimensional
chain. Although we do not explicitly consider electron CHT, which is
fermionic, the analogous forms of the heat transfer spectrum and
analogous properties of the relevant linear response quantities allow
for derivation of analogous rank-based and channel-based bounds. Our
findings have ramifications for the design of new nanoscale junctions
for efficient CHT.

\section{Rank-based bounds} \label{sec:rankbounds}

\begin{figure}[t!]
  \centering
  \includegraphics[width=0.95\columnwidth]{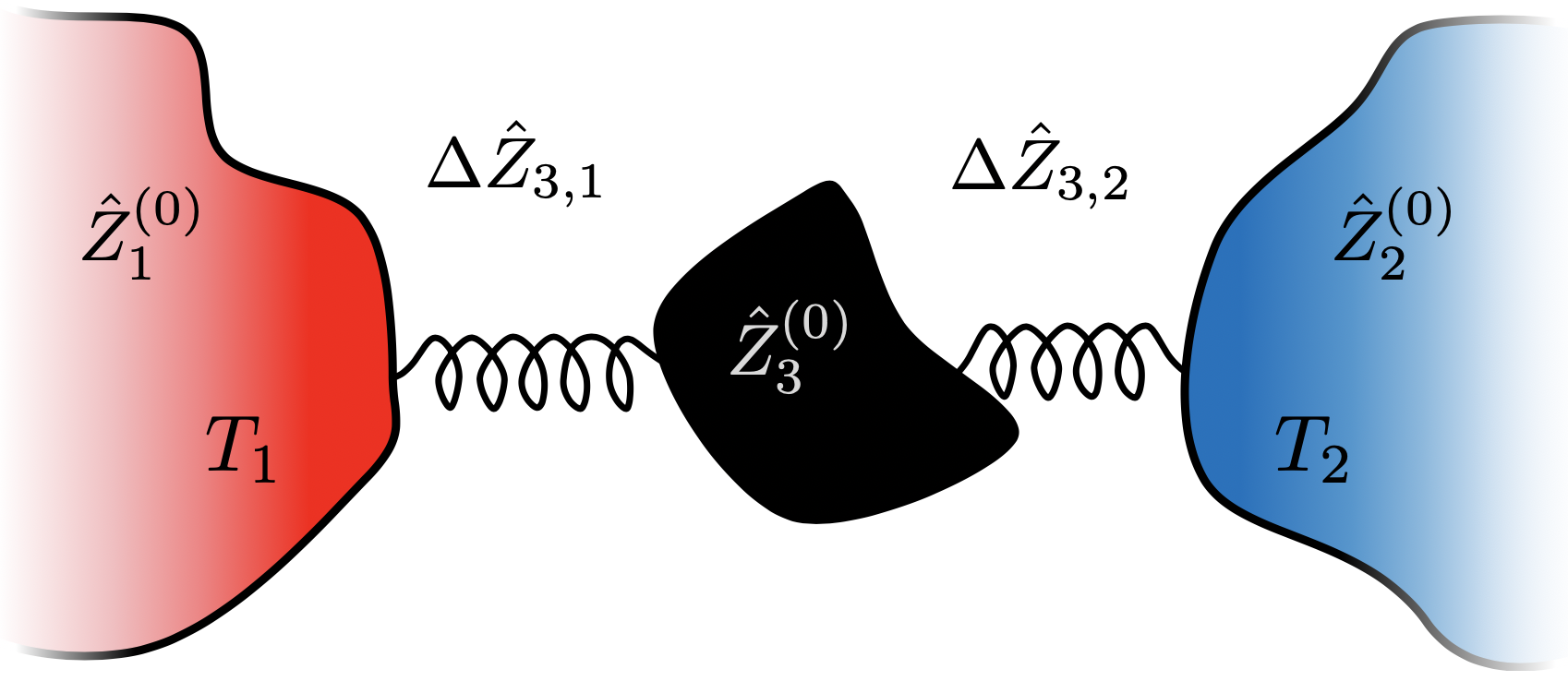}
  \caption{\textbf{Schematic system}. Two components, labeled 1 \& 2,
    with linear response functions $\hat{Z}^{(0)}_{1}$ and
    $\hat{Z}^{(0)}_{2}$ and maintained at temperatures $T_{1}$ and
    $T_{2}$, exchange energy by virtue of coupling to a third
    component, labeled 3 with linear response function
    $\hat{Z}^{(0)}_{3}$, through couplings $\Delta\hat{Z}_{3,1}$ and
    $\Delta\hat{Z}_{3,2}$. All relevant response quantities are
    assumed to be linear, reciprocal, causal, and passive.}
  \label{fig:schematic}
\end{figure}

We consider the system depicted schematically in~\figref{schematic},
consisting of two arbitrary components, labeled 1 and 2, exchanging
energy only via couplings to a third body 3. Physical embodiments
include RHT between any two bodies, as well as CHT for two bodies
connected only through a junction~\cite{VenkataramARXIV2020}. If each
component exhibits linear bosonic response with linear couplings, the
net heat transfer power may be written as
\begin{equation} \label{eq:generalP}
  P_{1 \to 2} = \int_{0}^{\infty} [\Pi(\omega, T_{2}) - \Pi(\omega,
  T_{1})]\Phi(\omega)~\frac{\mathrm{d}\omega}{2\pi}
\end{equation}
where $\Pi(\omega, T) = \hbar\omega/(e^{\hbar\omega/(k_{\mathrm{B}}
  T)} - 1)$ is the Planck function, while $\Phi(\omega)$ is the
dimensionless spectrum of energy transfer. The frequency domain
equations of motion defining the response of each component $n \in
\{1, 2, 3\}$ in isolation are given through the linear operators
$\hat{Z}^{(0)}_{n}$ and their inverses $\hat{Y}^{(0)}_{n} =
\hat{Z}^{(0)-1}_{n}$, and the linear coupling of each component $n \in
\{1, 2\}$ to component 3 is $\Delta \hat{Z}_{3,n}$. Reciprocity means
that at any frequency $\omega$, $\hat{Z}^{(0)}_{n} =
(\hat{Z}^{(0)}_{n})^{\top}$ must hold, though that may be
complex-valued, but we assume $\Delta \hat{Z}_{3,n} = (\Delta
\hat{Z}_{n,3})^{\top}$ to be real-valued. Furthermore, passivity means
that $\asym(\hat{Y}^{(0)}_{n}) = \Im(\hat{Y}^{(0)}_{n})$ for $n \in
\{1, 2, 3\}$ are all positive-semidefinite operators. Throughout this
paper, we assume that if the coupling induces an effective nonzero
$\Delta\hat{Z}_{nn}$ for each $n \in \{1, 2, 3\}$, that is included in
the definition of $\hat{Z}^{(0)}_{n}$. As examples, the relevant
material response functions $\hat{Z}^{(0)}_{n}$ for RHT will generally
include the effects of short-range Coulomb interactions, and the
electromagnetic couplings (charges) $\Delta\hat{Z}_{n,3}$ will only be
for explicitly long-range interactions; for phonon CHT, the specific
example of two 1D harmonic oscillators $n \in \{1, 2\}$ of masses
$m_{n}$ coupled to each other with strength $k_{1,2}$ and each to a
separate fixed wall with strength $k_{nn}$ has equations of motion
$(k_{1,1} + k_{1,2} - \omega^{2} m_{1})x_{1} - k_{1,2} x_{2} = F_{1}$
and $-k_{1,2} x_{1} + (k_{2,2} + k_{1,2} - \omega^{2} m_{2})x_{2} =
F_{2}$, so $\hat{Z}^{(0)}_{n} = k_{nn} + k_{1,2} - \omega^{2}
m_{n}$. Finally, throughout this paper, we assume that no degrees of
freedom (DOFs) in component 3 are simultaneously coupled to components
1 and 2, which is true of RHT when the material bodies do not overlap,
and can be assumed of CHT by expanding the definition of the central
junction to prohibit such overlaps; we do this for conceptual and
computational simplicity, though the formula for heat transfer and our
bounds can be generalized to incorporate other scenarios. Given this,
we may write the energy transfer spectrum at each $\omega$
as~\cite{VenkataramARXIV2020}
\begin{multline} \label{eq:Phiactual}
  \Phi = 4~\trace{\asym(\Delta \hat{Z}_{3,1} \hat{Y}^{(0)}_{1} \Delta
    \hat{Z}_{1,3}) \hat{Y}_{3}^{\dagger}\right. \times
    \\ \left.\asym(\Delta \hat{Z}_{3,2} \hat{Y}^{(0)}_{2} \Delta
    \hat{Z}_{2,3}) \hat{Y}_{3}},
\end{multline}
where we define $\hat{Y}_{3} = (\hat{Z}^{(0)}_{3} - \Delta
\hat{Z}_{3,1} \hat{Y}^{(0)}_{1} \Delta \hat{Z}_{1,3} - \Delta
\hat{Z}_{3,2} \hat{Y}^{(0)}_{2} \Delta \hat{Z}_{2,3})^{-1}$ as the
response of component 3 dressed by its couplings to components 1 and
2. For clarity, \appref{glossary} gives correspondences between these
abstract operators and concrete linear response quantities in the
context of phonon CHT; as examples, in the context of phonon CHT
between two leads across a junction, $\hat{Y}_{3}$ is the Green's
function of the junction dressed by the leads, and the operators
$\Delta\hat{Z}_{3,n} \asym(\hat{Y}^{(0)}_{n}) \Delta\hat{Z}_{n,3}$ for
$n \in \{1, 2\}$ are the imaginary parts of the self-energies of the
leads.

In the context of CHT, components 1 and 2 frequently correspond to
large leads, while component 3 corresponds to a much smaller
intermediate junction~\cite{BurklePRB2015, KlocknerPRB2016,
  KlocknerPRB2017A, KlocknerPRB2017B, KlocknerPRB2017C,
  KlocknerPRB2018}. As the operators in the trace in~\eqref{Phiactual}
can be cyclically rearranged, $\Phi$ can be written as the trace of a
Hermitian positive-semidefinite operator, and it has already been
shown~\cite{VenkataramARXIV2020} that its eigenvalues, representing
transmission values associated with individual conduction channels,
all lie in the range $[0, 1]$. Going further, we note that the
transmission eigenvalues are the squares of the singular values of the
operator $\asym(\Delta \hat{Z}_{3,2} \hat{Y}^{(0)}_{2} \Delta
\hat{Z}_{2,3})^{1/2} \hat{Y}_{3} \asym(\Delta \hat{Z}_{3,1}
\hat{Y}^{(0)}_{1} \Delta \hat{Z}_{1,3})^{1/2}$, lying in the same
range. The number of nonzero singular values is the rank of this
operator, which is bounded above by the rank of the various operators
being multiplied together, namely $\asym(\Delta \hat{Z}_{3,n}
\hat{Y}^{(0)}_{n} \Delta \hat{Z}_{n,3})^{1/2}$ for $n \in \{1, 2\}$,
along with $\hat{P}_{3(2)} \hat{Y}_{3} \hat{P}_{3(1)}$, where
$\hat{P}_{3(n)} = (\hat{P}_{3(n)})^{\top}$ is the orthogonal
projection into the subspace of component 3 coupled to component $n
\in \{1, 2\}$. Therefore, the number of nonzero transmission
eigenvalues is bounded above by the minimum of the number of DOFs of
component 3 coupled to component 1 versus 2, or the number of DOFs of
components 1 or 2 coupled to component 3 (in case the couplings are
not one-to-one per DOF in each component). However, even tighter
bounds on the number of nonzero transmission eigenvalues can be
derived with the following observation. Frequently, component 3 has a
narrow bottleneck that has even fewer DOFs than those coupling to
components 1 or 2, and if that bottleneck is not directly coupled to
components 1 or 2, then component 3 can be divided into subparts A, B,
and C, such that subparts A and B couple directly respectively to
components 1 and 2 and also to subpart C, while subpart C only couples
directly to subparts A and B. This means that $\hat{P}_{3(1)} =
\hat{P}_{3\mathrm{A}(1)}$ and $\hat{P}_{3(2)} =
\hat{P}_{3\mathrm{B}(2)}$, while we define the nonzero blocks of
$\Delta \hat{Z}_{1,3}$ and $\Delta \hat{Z}_{2,3}$ as $\Delta
\hat{Z}_{1,3\mathrm{A}}$ and $\Delta \hat{Z}_{2,3\mathrm{B}}$
respectively. Writing the operators in the space of component 3 in
block form in terms of the subparts, so that $\hat{Y}_{3}^{-1}$ is
given by
\begin{widetext}
\begin{equation*}
  \hat{Z}^{(0)}_{3} - \Delta \hat{Z}_{3,1} \hat{Y}^{(0)}_{1} \Delta
  \hat{Z}_{1,3} - \Delta \hat{Z}_{3,2} \hat{Y}^{(0)}_{2} \Delta
  \hat{Z}_{2,3} = \begin{bmatrix} \hat{Z}^{(0)}_{3\mathrm{A}} - \Delta
    \hat{Z}_{3\mathrm{A},1} \hat{Y}^{(0)}_{1} \Delta
    \hat{Z}_{1,3\mathrm{A}} & 0 & \Delta \hat{Z}_{3\mathrm{AC}} \\ 0 &
    \hat{Z}^{(0)}_{3\mathrm{B}} - \Delta \hat{Z}_{3\mathrm{B},2}
    \hat{Y}^{(0)}_{2} \Delta \hat{Z}_{2,3\mathrm{B}} & \Delta
    \hat{Z}_{3\mathrm{BC}} \\ \Delta \hat{Z}_{3\mathrm{CA}} & \Delta
    \hat{Z}_{3\mathrm{CB}} & \hat{Z}^{(0)}_{3\mathrm{C}}
  \end{bmatrix}
\end{equation*}
\end{widetext}
then we may perform the inversion blockwise to yield the B-A
off-diagonal subpart block, which is the only relevant nonzero block
for energy exchange in this scenario, as $\hat{P}_{3(2)} \hat{Y}_{3}
\hat{P}_{3(1)} = \hat{P}_{3\mathrm{B}(2)} (\hat{Z}^{(0)}_{3\mathrm{B}}
- \Delta \hat{Z}_{3\mathrm{B},2} \hat{Y}^{(0)}_{2} \Delta
\hat{Z}_{2,3\mathrm{B}})^{-1} \Delta \hat{Z}_{3\mathrm{BC}}
(\hat{Z}^{(0)}_{3\mathrm{C}} - \Delta \hat{Z}_{3\mathrm{CA}}
(\hat{Z}^{(0)}_{3\mathrm{A}} - \Delta \hat{Z}_{3\mathrm{A},1}
\hat{Y}^{(0)}_{1} \Delta \hat{Z}_{1,3\mathrm{A}})^{-1} \Delta
\hat{Z}_{3\mathrm{AC}} - \Delta \hat{Z}_{3\mathrm{CB}}
(\hat{Z}^{(0)}_{3\mathrm{B}} - \Delta \hat{Z}_{3\mathrm{B},2}
\hat{Y}^{(0)}_{2} \Delta \hat{Z}_{2,3\mathrm{B}})^{-1} \Delta
\hat{Z}_{3\mathrm{BC}})^{-1} \Delta \hat{Z}_{3\mathrm{CA}}
(\hat{Z}^{(0)}_{3\mathrm{A}} - \Delta \hat{Z}_{3\mathrm{A},1}
\hat{Y}^{(0)}_{1} \Delta \hat{Z}_{1,3\mathrm{A}})^{-1}
\hat{P}_{3\mathrm{A}(1)}$. In this expression, the middle operator is
nonzero only in the space of subpart C, so if subpart C has the fewest
DOFs, then it is the rank-limiting part. It follows that the number of
nonzero transmission eigenvalues for two leads exchanging energy via a
junction is bounded above by the number of DOFs in the smallest
bottleneck in the junction, validating prior
observations~\cite{CuevasPRL1998, BurklePRB2015, KlocknerPRB2016,
  KlocknerPRB2017A, KlocknerPRB2017B, KlocknerPRB2017C,
  KlocknerPRB2018}.

\section{Channel-based bounds} \label{sec:channelbounds}

In order to derive upper bounds on $\Phi$ (no longer assuming any
particular division of component 3 into subparts)
from~\eqref{Phiactual}, we rewrite $\Phi$ as follows. First, we
recognize that reciprocity of the relevant operators allows
replacement of the Hermitian adjoint ${}^{\dagger}$ by the complex
conjugate ${}^{\star}$, and in turn of $\asym$ by $\Im$. Second, we
assume that for each of the end components ($n \in \{1, 2\}$)
connected to the middle component ($3$), the operator $\Delta
\hat{Z}_{3,n} \hat{Y}^{(0)}_{n} \Delta \hat{Z}_{n,3}$ for a given $n$
is invertible within the space of DOFs of component 3 which are
coupled to that component $n$. This allows for rewriting 
\begin{multline*}
  \Im(\Delta \hat{Z}_{3,n} \hat{Y}^{(0)}_{n} \Delta \hat{Z}_{n,3}) =
  \Delta \hat{Z}_{3,n} \hat{Y}^{(0)\star}_{n} \Delta \hat{Z}_{n,3}
  \times \\ \Im((\Delta \hat{Z}_{3,n} \hat{Y}^{(0)\star}_{n} \Delta
  \hat{Z}_{n,3})^{-1}) \Delta \hat{Z}_{3,n} \hat{Y}^{(0)}_{n} \Delta
  \hat{Z}_{n,3}.
\end{multline*}
For notational convenience, in analogy with electromagnetic notation,
we denote $\hat{V}_{n} \equiv \Delta \hat{Z}_{3,n} \hat{Y}^{(0)}_{n}
\Delta \hat{Z}_{n,3}$; this denotes the response of component $n \in
\{1, 2\}$ evaluated in the space of DOFs of component 3 coupled to
component $n$ (multiplied by those coupling quantities), and in the
context of CHT, these are the self-energies of the leads $n$ coupled
to the junction (component 3). This therefore allows for writing
\begin{equation*}
  \Phi = 4~\trace{\hat{V}_{1} \Im(\hat{V}_{1}^{-1\star})
    \hat{V}_{1}^{\star} \hat{Y}_{3}^{\star} \hat{V}_{2}^{\star}
    \Im(\hat{V}_{2}^{-1\star}) \hat{V}_{2} \hat{Y}_{3}}
\end{equation*}
after rearranging the trace. It is also helpful at this point to use
the definition $\hat{Y}_{3} = (\hat{Z}^{(0)}_{3} - \hat{V}_{1} -
\hat{V}_{2})^{-1}$ to show that $\hat{V}_{2} \hat{Y}_{3} \hat{V}_{1} =
\hat{V}_{2} (\hat{1} - (\hat{1} - \hat{Y}^{(0)}_{3} \hat{V}_{1})^{-1}
\hat{Y}^{(0)}_{3} \hat{V}_{2})^{-1} \hat{Y}^{(0)}_{3} \hat{V}_{1}
(\hat{1} - \hat{Y}^{(0)}_{3} \hat{V}_{1})^{-1}$, as this will become
useful for the derivations of these bounds.

Having established these operator definitions and relations, our
derivation proceeds following the derivation of the RHT bounds
in~\citeasnoun{MoleskyPRB2020}. First, we establish and explain
generalizations of constraints on quantities like ``far-field
scattering'' relevant to RHT. Second, we apply a lemma by von
Neumann~\cite{MirskyMFM1975} (whose derivation is reproduced in
context in~\citeasnoun{MoleskyPRB2020}), showing that the largest real
positive value of the trace of a product of operators arises when
those operators share singular vectors and when the sets of fixed
singular values are arranged in consistent orders, to the maximization
of $\Phi$, explaining along the way how variation of the singular
values themselves is consistent with the conditions of the lemma. We
conclude the section by restating the bound for $\Phi$, which we term
$\Phi_{\mathrm{opt}}$, and by discussing its implications.

\subsection{Constraints on nonnegative far-field scattering}

In~\cite{VenkataramARXIV2020}, we established that in the context of
RHT, component 3 corresponds to the vacuum electromagnetic field, so
$\hat{Y}^{(0)}_{3}$ would be the vacuum Maxwell Green's function. We
extend this analogy in the other direction to derive constraints on
the linear response quantities relevant to this system of two
components coupled only via a third.

First, we define the relevant general equations of motion for these
operators in order to properly define what is meant by absorbed,
scattered, and extinguished power. In~\cite{VenkataramARXIV2020}, we
generally showed that they can be written as $\ket{x} = \ket{x^{(0)}}
+ \hat{Y}^{(0)} \ket{F}$ along with $\ket{F} = -\Delta\hat{Z} \ket{x}$
where there are nonzero generalized free displacements but no
generalized external forces. Here, we do the opposite in order to
describe powers in response to generalized forces, so the relevant
equations of motion are $\ket{x} = \hat{Y}^{(0)} \ket{F}$ and $\ket{F}
= \ket{F^{(0)}} - \Delta\hat{Z} \ket{x}$; for this particular
derivation, these are effectively related by the replacement
$\ket{F^{(0)}} \leftrightarrow \hat{Z}^{(0)} \ket{x}$. For this system
of two components labeled 1 \& 2 connected via a third labeled 3,
these quantities may be defined in block form as
\begin{multline}
  \ket{x} = \begin{bmatrix}
    \ket{x_{1}} \\
    \ket{x_{2}} \\
    \ket{x_{3}}
  \end{bmatrix},~
  \ket{F} = \begin{bmatrix}
    \ket{F_{1}} \\
    \ket{F_{2}} \\
    \ket{F_{3}}
  \end{bmatrix},~
  \ket{F^{(0)}} = \begin{bmatrix}
    \ket{F^{(0)}_{1}} \\
    \ket{F^{(0)}_{2}} \\
    \ket{F^{(0)}_{3}}
  \end{bmatrix} \\
  \hat{Z}^{(0)} = \begin{bmatrix}
    \hat{Z}^{(0)}_{1} & 0 & 0 \\
    0 & \hat{Z}^{(0)}_{2} & 0 \\
    0 & 0 & \hat{Z}^{(0)}_{3}
  \end{bmatrix},~
  \Delta\hat{Z} = \begin{bmatrix}
    0 & 0 & \Delta \hat{Z}_{1,3} \\
    0 & 0 & \Delta \hat{Z}_{2,3} \\
    \Delta \hat{Z}_{3,1} & \Delta \hat{Z}_{3,2} & 0
  \end{bmatrix}
\end{multline}
and for further convenience, we define the sub-groups,
\begin{multline}
  \ket{x_{\mathrm{A}}} = \begin{bmatrix}
    \ket{x_{1}} \\
    \ket{x_{2}}
  \end{bmatrix},~
  \ket{F_{\mathrm{A}}} = \begin{bmatrix}
    \ket{F_{1}} \\
    \ket{F_{2}}
  \end{bmatrix},~
  \ket{F^{(0)}_{\mathrm{A}}} = \begin{bmatrix}
    \ket{F^{(0)}_{1}} \\
    \ket{F^{(0)}_{2}}
  \end{bmatrix} \\
  \hat{Z}^{(0)}_{\mathrm{A}} = \begin{bmatrix}
    \hat{Z}^{(0)}_{1} & 0 \\
    0 & \hat{Z}^{(0)}_{2}
  \end{bmatrix},~
  \Delta\hat{Z}_{\mathrm{A},3} = \begin{bmatrix}
    \Delta \hat{Z}_{1,3} \\
    \Delta \hat{Z}_{2,3}
  \end{bmatrix}
\end{multline}
where the subscript ``A'' refers to the ``aggregate'' of the
components 1 \& 2 that are not directly coupled to each other (and
$\Delta\hat{Z}_{3,\mathrm{A}} =
(\Delta\hat{Z}_{\mathrm{A},3})^{\top}$).

Next, we define the notions of absorbed, scattered, and extinguished
power in this system. For RHT, it is simple to see that far-field
scattering involves the transfer of energy to component 3, namely the
vacuum electromagnetic field, while absorption involves the transfer
of energy to components 1 or 2~\cite{MillerOE2016}. We generalize this
as follows. We assume that all external forces are only in components
1 or 2, so $\ket{F^{(0)}_{3}} = 0$, but $\ket{F^{(0)}_{\mathrm{A}}}
\neq 0$. We also define the orthogonal projection operators
$\hat{P}_{n}$ which project onto the subspaces of DOFs of component $n
\in \{1, 2, 3\}$; these are orthogonal to each other, so we also
define the ``aggregate'' projection $\hat{P}_{\mathrm{A}} =
\hat{P}_{1} + \hat{P}_{2}$. Similarly, we define the ``aggregate''
response operator $\hat{V}_{\mathrm{A}} = \hat{V}_{1} + \hat{V}_{2} =
\Delta\hat{Z}_{3,\mathrm{A}} \hat{Y}^{(0)}_{\mathrm{A}}
\Delta\hat{Z}_{\mathrm{A},3}$, but we do not yet assume that the DOFs
of component 3 that couple to each of the other components exist in
orthogonal subspaces. From these definitions, one finds that absorbed
power $\Phi_{\mathrm{abs}} = \frac{\omega}{2} \Im(\bracket{F,
  \hat{P}_{\mathrm{A}} x})$ refers to the energy dumped by the
external forces into components 1 \& 2, extinguished power
$\Phi_{\mathrm{ext}} = \frac{\omega}{2} \Im(\bracket{F^{(0)},
  \hat{P}_{\mathrm{A}} x})$ is the energy dumped by the external
forces into the entire system, and scattered power
$\Phi_{\mathrm{sca}} = \Phi_{\mathrm{ext}} - \Phi_{\mathrm{abs}}$ is
simply the difference of the two.

At this point, we may now generally compute these power quantities for
a general external force $\ket{F^{(0)}_{\mathrm{A}}}$. The equations
of motion may be written as
\begin{equation} \label{eq:EOMAand3}
  \begin{split}
    \hat{Z}^{(0)}_{\mathrm{A}} \ket{x_{\mathrm{A}}} +
    \Delta\hat{Z}_{\mathrm{A},3} \ket{x_{3}} &=
    \ket{F^{(0)}_{\mathrm{A}}} \\ \Delta\hat{Z}_{3,\mathrm{A}}
    \ket{x_{\mathrm{A}}} + \hat{Z}^{(0)}_{3} \ket{x_{3}} &= 0
  \end{split}
\end{equation}
whose formal solution may be written as
\begin{equation}
  \begin{split}
    \ket{x_{\mathrm{A}}} &= \hat{Y}^{(0)}_{\mathrm{A}} (\hat{1} +
    \Delta\hat{Z}_{\mathrm{A},3} \hat{Y}^{(0)}_{3} (\hat{1} -
    \hat{V}_{\mathrm{A}} \hat{Y}^{(0)}_{3})^{-1}
    \Delta\hat{Z}_{3,\mathrm{A}} \hat{Y}^{(0)}_{\mathrm{A}})
    \ket{F^{(0)}_{\mathrm{A}}} \\ \ket{x_{3}} &= -\hat{Y}^{(0)}_{3}
    (\hat{1} - \hat{V}_{\mathrm{A}} \hat{Y}^{(0)}_{3})^{-1}
    \Delta\hat{Z}_{3,\mathrm{A}} \hat{Y}^{(0)}_{\mathrm{A}}
    \ket{F^{(0)}_{\mathrm{A}}}
  \end{split}
\end{equation}
in terms of $\ket{F^{(0)}_{\mathrm{A}}}$. From this, we write the
absorbed power $\Phi_{\mathrm{abs}} = \frac{\omega}{2} \Im(\bracket{F,
  \hat{P}_{\mathrm{A}} x})$ (after relevant operator manipulations) as
\begin{multline}
  \frac{2}{\omega} \Phi_{\mathrm{abs}} =
  \langle F^{(0)}_{\mathrm{A}}, \hat{Y}^{(0)\star}_{\mathrm{A}}
    \Delta\hat{Z}_{\mathrm{A},3} (\hat{1} - \hat{Y}^{(0)\star}_{3}
    \hat{V}_{\mathrm{A}}^{\star})^{-1} \hat{Y}^{(0)\star}_{3} \times \\
    \Im(\hat{V}_{\mathrm{A}}) \hat{Y}^{(0)}_{3} (\hat{1} -
    \hat{V}_{\mathrm{A}} \hat{Y}^{(0)}_{3})^{-1}
    \Delta\hat{Z}_{3,\mathrm{A}} \hat{Y}^{(0)}_{\mathrm{A}}
    F^{(0)}_{\mathrm{A}} \rangle
\end{multline}
upon using the real-valued nature of $\Delta\hat{Z}_{\mathrm{A},3}$
and its transpose, and the fact that for operators $\hat{A}$ and
$\hat{B}$, $\asym(\hat{A}^{\dagger} \hat{B} \hat{A}) =
\hat{A}^{\dagger} \asym(\hat{B}) \hat{A}$ (with a similar statement
holding for reciprocal operators with the complex conjugate and the
imaginary part); passivity means that $\Im(\hat{V}_{n})$ is
positive-semidefinite for $n \in \{1, 2\}$, and so is
$\Im(\hat{V}_{\mathrm{A}})$ in turn, guaranteeing that
$\Phi_{\mathrm{abs}} \geq 0$ for any
$\ket{F^{(0)}_{\mathrm{A}}}$. Likewise, we write the extinguished
power $\Phi_{\mathrm{ext}} = \frac{\omega}{2} \Im(\bracket{F^{(0)},
  \hat{P}_{\mathrm{A}} x})$ (after relevant operator manipulations) as
\begin{multline}
  \frac{2}{\omega} \Phi_{\mathrm{ext}} =
  \langle F^{(0)}_{\mathrm{A}}, \hat{Y}^{(0)\star}_{\mathrm{A}}
    \Delta\hat{Z}_{\mathrm{A},3} \times \\ \Im((\hat{1} - \hat{Y}^{(0)}_{3}
    \hat{V}_{\mathrm{A}})^{-1} \hat{Y}^{(0)}_{3})
    \Delta\hat{Z}_{3,\mathrm{A}} \hat{Y}^{(0)}_{\mathrm{A}}
    F^{(0)}_{\mathrm{A}} \rangle
\end{multline}
for which it can be shown that $\Phi_{\mathrm{ext}} \geq 0$ for all
$\ket{F^{(0)}_{\mathrm{A}}}$ as follows: performing all inverses in
the space corresponding to DOFs of component 3, $\Im((\hat{1} -
\hat{Y}^{(0)}_{3} \hat{V}_{\mathrm{A}})^{-1} \hat{Y}^{(0)}_{3}) =
\Im((\hat{Z}^{(0)}_{3} - \hat{V}_{\mathrm{A}})^{-1}) =
(\hat{Z}^{(0)}_{3} - \hat{V}_{\mathrm{A}})^{-1\star}
\Im(\hat{Z}^{(0)\star}_{3} - \hat{V}_{\mathrm{A}}^{\star})
(\hat{Z}^{(0)}_{3} - \hat{V}_{\mathrm{A}})^{-1}$, so this is
positive-semidefinite if $\Im(\hat{Z}^{(0)\star}_{3} -
\hat{V}_{\mathrm{A}}^{\star})$ is positive-semidefinite, which is true
as passivity means each term, namely $\Im(\hat{Z}^{((0)\star)}_{3})$
and $-\Im(\hat{V}_{\mathrm{A}}^{\star})$, is positive-semidefinite.

In order to show that the difference between the extinguished and
absorbed powers is properly a scattered power and is nonnegative, we
must show that it is equal to the energy dumped into component 3. To
do this, we may rewrite the solution to~\eqref{EOMAand3} in a fully
equivalent way as
\begin{equation}
  \begin{split}
    \ket{x_{\mathrm{A}}} &= (\hat{1} - \hat{Y}^{(0)}_{\mathrm{A}}
    \Delta\hat{Z}_{\mathrm{A},3} \hat{Y}^{(0)}_{3}
    \Delta\hat{Z}_{3,\mathrm{A}})^{-1} \hat{Y}^{(0)}_{\mathrm{A}}
    \ket{F^{(0)}_{\mathrm{A}}} \\
    \ket{x_{3}} &= -\hat{Y}^{(0)}_{3} \Delta\hat{Z}_{3,\mathrm{A}}
    (\hat{1} - \hat{Y}^{(0)}_{\mathrm{A}} \Delta\hat{Z}_{\mathrm{A},3}
    \hat{Y}^{(0)}_{3} \Delta\hat{Z}_{3,\mathrm{A}})^{-1}
    \hat{Y}^{(0)}_{\mathrm{A}} \ket{F^{(0)}_{\mathrm{A}}}
  \end{split}
\end{equation}
and then write the absorbed power in component 3 (ignoring the
prefactor $\omega/2$) as $\Im(\bracket{F, \hat{P}_{3} x}) =
-\Im(\bracket{\Delta \hat{Z}_{3,\mathrm{A}} x_{\mathrm{A}},
  x_{3}})$. This in turn is written (after relevant operator
manipulations) as
\begin{multline}
  -\Im(\bracket{\Delta \hat{Z}_{3,\mathrm{A}} x_{\mathrm{A}}, x_{3}})
  = \\ \langle F^{(0)}_{\mathrm{A}}, \hat{Y}^{(0)\star}_{\mathrm{A}}
  \Delta\hat{Z}_{\mathrm{A},3} (\hat{1} - \hat{Y}^{(0)\star}_{3}
  \hat{V}^{\star}_{\mathrm{A}})^{-1} \Im(\hat{Y}^{(0)}_{3}) \times
  \\ (\hat{1} - \hat{V}_{\mathrm{A}} \hat{Y}^{(0)}_{3})^{-1}
  \Delta\hat{Z}_{3,\mathrm{A}} \hat{Y}^{(0)}_{\mathrm{A}}
  F^{(0)}_{\mathrm{A}} \rangle
\end{multline}
and this is indeed equal to $\Phi_{\mathrm{ext}} -
\Phi_{\mathrm{abs}}$, as the equality $(\hat{1} -
\hat{Y}^{(0)\star}_{3} \hat{V}^{\star}_{\mathrm{A}})^{-1}
\Im(\hat{Y}^{(0)}_{3}) (\hat{1} - \hat{V}_{\mathrm{A}}
\hat{Y}^{(0)}_{3})^{-1} = \Im((\hat{1} - \hat{Y}^{(0)}_{3}
\hat{V}_{\mathrm{A}})^{-1} \hat{Y}^{(0)}_{3}) - (\hat{1} -
\hat{Y}^{(0)\star}_{3} \hat{V}_{\mathrm{A}}^{\star})^{-1}
\hat{Y}^{(0)\star}_{3} \Im(\hat{V}_{\mathrm{A}}) \hat{Y}^{(0)}_{3}
(\hat{1} - \hat{V}_{\mathrm{A}} \hat{Y}^{(0)}_{3})^{-1}$ follows from
the aforementioned operator identity $\Im((\hat{1} - \hat{Y}^{(0)}_{3}
\hat{V}_{\mathrm{A}})^{-1} \hat{Y}^{(0)}_{3}) = \Im((\hat{Z}^{(0)}_{3}
- \hat{V}_{\mathrm{A}})^{-1}) = (\hat{Z}^{(0)}_{3} -
\hat{V}_{\mathrm{A}})^{-1\star} \Im(\hat{Z}^{(0)\star}_{3} -
\hat{V}_{\mathrm{A}}^{\star}) (\hat{Z}^{(0)}_{3} -
\hat{V}_{\mathrm{A}})^{-1}$. Thus, the energy dumped in component 3 is
indeed a scattered power, and passivity, namely the operator
$\Im(\hat{Y}^{(0)}_{3})$ being positive-semidefinite, makes it
nonnegative. This in turn requires that the equivalent operator
$\Im((\hat{1} - \hat{Y}^{(0)}_{3} \hat{V}_{\mathrm{A}})^{-1}
\hat{Y}^{(0)}_{3}) - (\hat{1} - \hat{Y}^{(0)\star}_{3}
\hat{V}_{\mathrm{A}}^{\star})^{-1} \hat{Y}^{(0)\star}_{3}
\Im(\hat{V}_{\mathrm{A}}) \hat{Y}^{(0)}_{3} (\hat{1} -
\hat{V}_{\mathrm{A}} \hat{Y}^{(0)}_{3})^{-1}$ be
positive-semidefinite. Furthermore, the existence of nontrivial
scattered power requires that $\Im(\hat{Y}^{(0)}_{3}) \neq 0$, which
requires some form of dissipation in component 3; in RHT, this is
automatically satisfied by far-field radiation encoded in the vacuum
Maxwell Green's function, but for other forms of heat transfer, e.g.,
CHT through an intermediate small junction (functioning as component
3), dissipation must be explicitly introduced. This is required for
self-consistency, but as will become clear, $\Im(\hat{Y}^{(0)}_{3})$
does not appear in other forms of the constraint requiring nonnegative
scattered power, and ultimately does not affect bounds
on~\eqref{Phiactual}.

We now explore the consequences of this constraint for each component
$n \in \{1, 2\}$ coupled to component 3, and particularly wish to cast
these constraints in terms of the DOFs of component 3 that are coupled
to each of the other components, thereby involving the operators
$\hat{V}_{n}$. With this in mind, we first consider the implications
of nonnegative far-field scattering in the absence of component 2, so
component 3 is only coupled to component 1, and compute the absorbed
power in component 3 (i.e. the scattered power) due to the external
force $\ket{F^{(0)}_{1}}$. The exact same steps as above can be
followed under the notational replacement $\mathrm{A} \to 1$. Starting
from the condition that $\hat{Y}^{(0)\star}_{1} \Delta\hat{Z}_{1,3}
(\hat{1} - \hat{Y}^{(0)\star}_{3} \hat{V}^{\star}_{1})^{-1}
\Im(\hat{Y}^{(0)}_{3}) (\hat{1} - \hat{V}_{1} \hat{Y}^{(0)}_{3})^{-1}
\Delta\hat{Z}_{3,1} \hat{Y}^{(0)}_{1}$ be positive-semidefinite, we
multiply on the left by $\Delta\hat{Z}_{3,1}$ and on the right by
$\Delta\hat{Z}_{1,3}$, which does not affect this condition. We then
define the operator $\hat{T}_{1} = (\hat{1} - \hat{V}_{1}
\hat{Y}^{(0)}_{3})^{-1} \hat{V}_{1}$ (not to be confused with the
temperature $T_{1}$), which is reciprocal, and assume that it and
$\hat{V}_{1}$ are invertible within the space of the subset of DOFs of
component 3 coupled to component 1; this operator, known as the
T-operator in the context of electromagnetic scattering theory (or the
T-matrix in the context of electron or phonon scattering theories),
describes the response of the DOFs of component 3 coupled to component
1 dressed by the propagation of forces through the whole of component
3 in isolation. Using this, we define the orthogonal projection
operator onto that space as $\hat{P}(\hat{V}_{1})$ in order to say
that $\hat{P}(\hat{V}_{1}) \Im(\hat{Y}^{(0)}_{3}) \hat{P}(\hat{V}_{1})
= \Im(\hat{T}_{1}^{-1\star} - \hat{V}_{1}^{-1\star})$. Plugging this
projector yields the condition that scattering is nonnegative when the
operator $\Im(\hat{T}_{1}) - \hat{T}_{1}^{\star}
\Im(\hat{V}_{1}^{-1\star}) \hat{T}_{1}$ is positive-semidefinite. As
this yields a nonnegative quadratic form for every
$\ket{F^{(0)}_{1}}$, that will also be true for every
$\ket{F^{(0)\star}_{1}}$, so that in conjunction with reciprocity, one
finds that $\Im(\hat{T}_{1}) - \hat{T}_{1} \Im(\hat{V}_{1}^{-1\star})
\hat{T}_{1}^{\star}$ must also be positive-semidefinite.

Next, we consider the implications of nonnegative far-field scattering
for the full system, in which all three components are present with
components 1 \& 2 only coupled to component 3. At this point, we must
impose the assumption that no DOFs in component 3 are simultaneously
coupled to components 1 \& 2. (Relaxation of this assumption and
associated computational aspects are covered
in~\appref{alternatebounds}.) This means the two operators
$\hat{V}_{n}$ are supported in disjoint (orthogonal) subspaces, so
those quantities are in fact separable. Undoing the replacement
$\mathrm{A} \to 1$ above means that we can write $\hat{T}_{\mathrm{A}}
= (\hat{1} - \hat{V}_{\mathrm{A}} \hat{Y}^{(0)}_{3})^{-1}
\hat{V}_{\mathrm{A}}$, and analogously $\hat{P}(\hat{V}_{\mathrm{A}})
\Im(\hat{Y}^{(0)}_{3}) \hat{P}(\hat{V}_{\mathrm{A}}) =
\Im(\hat{T}_{\mathrm{A}}^{-1\star} - \hat{V}_{\mathrm{A}}^{-1\star})$,
so the condition for nonnegative far-field scattering (i.e. energy
dumped into component 3) is that $\Im(\hat{T}_{\mathrm{A}}) -
\hat{T}_{\mathrm{A}}^{\star} \Im(\hat{V}_{\mathrm{A}}^{-1\star})
\hat{T}_{\mathrm{A}}$ is positive-semidefinite. This analysis is made
more convenient by writing the relevant operators in block form as
\begin{equation}
  \begin{split}
    \hat{V}_{\mathrm{A}} &= \begin{bmatrix}
      \hat{V}_{1} & 0 \\
      0 & \hat{V}_{2}
    \end{bmatrix} \\
    \hat{T}_{\mathrm{A}}^{-1} &= \begin{bmatrix}
      \hat{V}_{1}^{-1} - \hat{P}(\hat{V}_{1}) \hat{Y}^{(0)}_{3}
      \hat{P}(\hat{V}_{1}) & -\hat{P}(\hat{V}_{1}) \hat{Y}^{(0)}_{3}
      \hat{P}(\hat{V}_{2}) \\
      -\hat{P}(\hat{V}_{2}) \hat{Y}^{(0)}_{3} \hat{P}(\hat{V}_{1}) &
      \hat{V}_{2}^{-1} - \hat{P}(\hat{V}_{2}) \hat{Y}^{(0)}_{3}
      \hat{P}(\hat{V}_{2})
    \end{bmatrix}
  \end{split}
\end{equation}
where each block represents a projection onto the space of the subset
of DOFs of component 3 coupled to each of the other components, and
where $\hat{V}_{1}$ and $\hat{V}_{2}$ are assumed to be invertible in
those spaces. The lower-right block of $\Im(\hat{T}_{\mathrm{A}}) -
\hat{T}_{\mathrm{A}}^{\star} \Im(\hat{V}_{\mathrm{A}}^{-1\star})
\hat{T}_{\mathrm{A}}$ may then be evaluated (upon further operator
manipulations) as $\Im(\hat{T}_{2,2}) - \hat{T}_{2,2}^{\star}
(\Im(\hat{V}_{2}^{-1\star}) + \hat{Y}^{(0)\star}_{3}
\hat{T}_{1}^{\star} \Im(\hat{V}_{1}^{-1\star}) \hat{T}_{1}
\hat{Y}^{(0)}_{3}) \hat{T}_{2,2}$, and this operator must be
positive-semidefinite, having defined $\hat{T}_{2,2} =
\hat{P}(\hat{V}_{2}) \hat{T}_{\mathrm{A}} \hat{P}(\hat{V}_{2}) =
(\hat{V}_{2}^{-1} - \hat{P}(\hat{V}_{2}) (\hat{Y}^{(0)}_{3} +
\hat{Y}^{(0)}_{3} \hat{T}_{1}
\hat{Y}^{(0)}_{3})\hat{P}(\hat{V}_{2}))^{-1}$ as the effective
response of the subset of DOFs of component 2 dressed by the
propagation of force through the whole of component 3 in the presence
of component 1. Reciprocity again means that the transpose, namely
$\Im(\hat{T}_{2,2}) - \hat{T}_{2,2} (\Im(\hat{V}_{2}^{-1\star}) +
\hat{Y}^{(0)}_{3} \hat{T}_{1} \Im(\hat{V}_{1}^{-1\star})
\hat{T}_{1}^{\star} \hat{Y}^{(0)\star}_{3}) \hat{T}_{2,2}^{\star}$,
must also be positive-semidefinite.

To summarize, nonnegative far-field scattering from component 1 when
component 3 is coupled only to it means that
\begin{equation} \label{eq:possca130}
  \bracket{u_{3}, \left[\Im(\hat{T}_{1}) - \hat{T}_{1}^{\star}
      \Im(\hat{V}_{1}^{-1\star}) \hat{T}_{1}\right] u_{3}} \geq 0
\end{equation}
must hold for any vector $\ket{u_{3}}$ in the space of component 3,
and the same must be true of the transpose of the relevant overall
operator. Likewise, nonnegative far-field scattering from component 2
when component 3 is coupled to both it and component 1 means that
\begin{multline} \label{eq:possca231}
  \langle u_{3}, \left[\Im(\hat{T}_{2,2}) - \hat{T}_{2,2}^{\star}
    \left(\Im(\hat{V}_{2}^{-1\star}) \right.\right. \\ \left.\left. +
    \hat{Y}^{(0)\star}_{3} \hat{T}_{1}^{\star}
    \Im(\hat{V}_{1}^{-1\star}) \hat{T}_{1} \hat{Y}^{(0)}_{3}\right)
    \hat{T}_{2,2}\right] u_{3} \rangle \geq 0
\end{multline}
must hold for any vector $\ket{u_{3}}$ in the space of component 3,
and the same must be true of the transpose of the overall operator.

\subsection{Optimization of singular values}
In order to optimize $\Phi$, we must rewrite it in a form that
explicitly depends on $\hat{T}_{1}$ and $\hat{T}_{2,2}$. To do this,
we start by returning to the operator identity $\hat{V}_{2}
\hat{Y}_{3} \hat{V}_{1} = \hat{V}_{2} (\hat{1} - (\hat{1} -
\hat{Y}^{(0)}_{3} \hat{V}_{1})^{-1} \hat{Y}^{(0)}_{3}
\hat{V}_{2})^{-1} \hat{Y}^{(0)}_{3} \hat{V}_{1} (\hat{1} -
\hat{Y}^{(0)}_{3} \hat{V}_{1})^{-1}$ presented at the beginning of
this section, and use the definitions of $\hat{T}_{1}$ and
$\hat{T}_{2,2}$ to show (upon manipulation of relevant operators) that
$\hat{V}_{2} \hat{Y}_{3} \hat{V}_{1} = \hat{T}_{2,2} \hat{Y}^{(0)}_{3}
\hat{T}_{1}$. This allows for immediately rewriting
\begin{equation}
  \Phi = 4~\trace{\Im(\hat{V}_{1}^{-1\star}) \hat{T}_{1}^{\star}
    \hat{Y}^{(0)\star}_{3} \hat{T}_{2,2}^{\star}
    \Im(\hat{V}_{2}^{-1\star}) \hat{T}_{2,2} \hat{Y}^{(0)}_{3}
    \hat{T}_{1}}
\end{equation}
and it is this form of $\Phi$ that shall be used to derive an upper
bound. In particular, the operators $\hat{V}_{1}$, $\hat{V}_{2}$, and
$\hat{Y}^{(0)}_{3}$ effectively describing the response of each
component in isolation will be taken to be fixed, while singular value
decompositions of $\hat{T}_{1}$ and $\hat{T}_{2,2}$ will be performed
in order to optimize the singular values to produce a bound on
$\Phi$. To do this, we further rewrite $\Phi = 4~\trace{\hat{A}
  \hat{B}_{2}^{\star} \hat{B}_{2}}$ upon defining $\hat{B}_{2} =
\Im(\hat{V}_{2}^{-1\star})^{1/2} \hat{T}_{2,2}
\Im(\hat{V}_{2}^{-1\star})^{1/2}$ and the Hermitian operator $\hat{A}
= \Im(\hat{V}_{2}^{-1\star})^{-1/2} \hat{Y}^{(0)}_{3} \hat{T}_{1}
\Im(\hat{V}_{1}^{-1\star}) \hat{T}_{1}^{\star} \hat{Y}^{(0)\star}_{3}
\Im(\hat{V}_{2}^{-1\star})^{-1/2}$.

This definition is convenient for the following reason. We wish to
vary the singular values of $\hat{T}_{1}$ and $\hat{T}_{2,2}$ to find
an upper bound for $\Phi$, yet the lemma by von
Neumann~\cite{MirskyMFM1975}, which states that the trace of a product
of operators has a maximum real nonnegative value when the singular
vectors among the operators are shared, requires the singular values
to be fixed in a consistent order; in our case, the constraints on the
various operators may depend on different mixtures of singular values
and singular vectors. However, our derivations are consistent with
this lemma thanks to the definitions of the operators $\hat{A}$ and
$\hat{B}_{2}$ above: in the definition of $\hat{A}$, only the singular
values of $\hat{T}_{1}$ may be varied, but the constraints on those
singular values are independent of constraints on the singular values
and vectors of other relevant operators. In particular, we may use the
reciprocity of $\hat{T}_{1}$ to write the singular value decomposition
$\hat{T}_{1} = \sum_{\mu} \tau_{(1)\mu} \ket{a_{\mu}}
\bra{a_{\mu}^{\star}}$, where $\bracket{a_{\mu}, a_{\nu}} =
\delta_{\mu\nu}$. Thus, if the singular values $\tau_{1(\mu)}$ are
appropriately set, the derivations are consistent with the
lemma~\cite{MirskyMFM1975}. We choose to write the singular value
decomposition $\hat{A} = \sum_{\nu} \alpha_{\nu} \ket{b_{\nu}}
\bra{b_{\nu}}$, given that $\hat{A}$ is a Hermitian
positive-semidefinite operator, and the constraint
in~\eqref{possca231} in transposed form is that $\Im(\hat{B}_{2}) -
\hat{B}_{2} (\hat{1} + \hat{A})\hat{B}_{2}^{\star}$ must be
positive-semidefinite, so we bound $\Phi = 4~\trace{\hat{B}_{2}
  \hat{A} \hat{B}_{2}^{\star}} \leq 4~\trace{\Im(\hat{B}_{2}) -
  \hat{B}_{2} \hat{B}_{2}^{\star}}$. From this, we can immediately see
that the right-hand side is maximized if $\hat{B}_{2} =
\im\Im(\hat{B}_{2})$ is purely anti-Hermitian (and still reciprocal,
as $\Im(\hat{V}_{2}^{-1\star})^{1/2}$ is not only Hermitian for a
passive system but is real-symmetric due to reciprocity), as any
nontrivial Hermitian part increases the magnitude of the negative
contribution on the right-hand side relative to the positive
contribution. Moreover, while the right-hand side is
basis-independent, it can be evaluated in the basis of singular
vectors $\{\ket{b_{\nu}}\}$ of $\hat{A}$, so the overall sum (trace)
is guaranteed to be maximized when each individual contribution is
maximized. The constraint that $\Im(\hat{B}_{2}) - \hat{B}_{2}
(\hat{1} + \hat{A})\hat{B}_{2}^{\star}$ must be positive-semidefinite
can be evaluated for a particular $\ket{b_{\mu}}$ as $\sum_{\nu}
\alpha_{\nu} |\bracket{b_{\nu}, \hat{T}_{2,2} b_{\mu}}|^{2} \leq
\bracket{b_{\mu}, \Im(\hat{T}_{2,2}) b_{\mu}} - \sum_{\nu}
|\bracket{b_{\nu}, \hat{T}_{2,2} b_{\mu}}|^{2}$, and so the right-hand
side is maximized for each channel $\mu$ if $\hat{T}_{2,2}$ has
$\{\ket{b_{\nu}}\}$ as its right singular vectors. Thus, the lemma by
von Neumann~\cite{MirskyMFM1975} is indeed applicable, and reciprocity
allows us to write the singular value decomposition $\hat{T}_{2,2} =
\sum_{\nu} \tau_{(2,2)\nu} \ket{b_{\nu}^{\star}} \bra{b_{\nu}}$, where
$\bracket{b_{\mu}, b_{\nu}} = \delta_{\mu\nu}$ and $\bracket{b_{\nu},
  b_{\nu}^{\star}} = \im$ for each channel $\nu$. In this basis of
singular vectors, we may write $\Phi \leq 4\sum_{\nu} \alpha_{\nu}
\tau_{(2,2)\nu}^{2}$, and the constraint can be written as
$\tau_{(2,2)\nu} - \tau_{(2,2)\nu}^{2} (1 + \alpha_{\nu}) \geq 0$, so
$\tau_{(2,2)\nu} \leq (1 + \alpha_{\nu})^{-1}$. For each
$\alpha_{\nu}$, the bound is saturated when the inequality on
$\tau_{(2,2)\nu}$ is saturated, so we may write $\Phi \leq 4\sum_{\nu}
\alpha_{\nu}/(1 + \alpha_{\nu})^{2}$ and then optimize each
$\alpha_{\nu}$ to maximize that bound.

At this point, we further rewrite $\hat{A} =
\Im(\hat{V}_{2}^{-1\star})^{-1/2} \hat{Y}^{(0)}_{3}
\Im(\hat{V}_{1}^{-1\star})^{-1/2} \hat{B}_{1} \times
\hat{B}_{1}^{\star} \Im(\hat{V}_{1}^{-1\star})^{-1/2}
\hat{Y}^{(0)\star}_{3} \Im(\hat{V}_{2}^{-1\star})^{-1/2}$ in terms of
$\hat{B}_{1} = \Im(\hat{V}_{1}^{-1\star})^{1/2} \hat{T}_{1}
\Im(\hat{V}_{1}^{-1\star})^{1/2}$. Because $\hat{T}_{1}$ has singular
values that may be freely chosen subject to constraints on nonnegative
scattering that are independent of the singular vectors, the largest
range of singular values $\alpha_{\nu}$ of $\hat{A}$ allowing for the
largest possible maximal value of the upper bound is thus made
available when $\hat{B}_{1}$ shares singular vectors with
$\Im(\hat{V}_{2}^{-1\star})^{-1/2} \hat{Y}^{(0)}_{3}
\Im(\hat{V}_{1}^{-1\star})^{-1/2}$, and the structure of $\hat{A}$
would then imply that $\Im(\hat{V}_{2}^{-1\star})^{-1/2}
\hat{Y}^{(0)}_{3} \Im(\hat{V}_{1}^{-1\star})^{-1/2}$ has
$\{\bra{b_{\nu}}\}$ as its left singular vectors too, while
$\Im(\hat{V}_{2}^{-1\star})^{-1/2} \hat{Y}^{(0)}_{3}
\Im(\hat{V}_{1}^{-1\star})^{-1/2}$ would have $\{\ket{a_{\mu}}\}$ as
its right singular vectors. Thus, we rewrite $\alpha_{\mu} =
(\kappa_{\mu} \eta_{\mu})^{2}$, where $\kappa_{\mu}$ are the (fixed)
singular values of $\Im(\hat{V}_{2}^{-1\star})^{-1/2}
\hat{Y}^{(0)}_{3} \Im(\hat{V}_{1}^{-1\star})^{-1/2}$ while
$\eta_{\mu}$ are the (variable) singular values of $\hat{B}_{1}$. We
thus rewrite $\Phi \leq 4\sum_{\mu} (\kappa_{\mu} \eta_{\mu})^{2} / (1
+ (\kappa_{\mu} \eta_{\mu})^{2})^{2}$. The contribution for each
channel $\mu$ is maximized at $\kappa_{\mu} \eta_{\mu} = 1$,
recovering the Landauer transmission bound of unity per channel for
those channels. However, this must be consistent with the constraint
in~\eqref{possca130} in its transposed form, which can be written as
the constraint that $\Im(\hat{B}_{1}) - \hat{B}_{1}
\hat{B}_{1}^{\star}$ be positive-semidefinite. Using similar arguments
as above, the optimal $\hat{B}_{1}$ should be purely anti-Hermitian
for the constraints on the singular values $\eta_{\mu}$ to be loosest,
so this is equivalent to the constraint that $\eta_{\mu} \leq 1$ for
each channel $\mu$. Therefore, if $\kappa_{\mu} \geq 1$ for a given
channel $\mu$, then we choose $\eta_{\mu} = 1/\kappa_{\mu}$ and
recover the Landauer transmission bound of unity for that
channel. Otherwise, if $\kappa_{\mu} < 1$, we must use the saturation
condition $\eta_{\mu} = 1$, for which the contribution to channel
$\mu$ is $\frac{4\kappa_{\mu}^{2}}{(1 + \kappa_{\mu}^{2})^{2}}$.

This bound can be succinctly written as
\begin{equation}
  \Phi \leq \sum_{\mu} \left[\Theta(\kappa_{\mu} - 1) +
    \frac{4\kappa_{\mu}^{2}}{(1 + \kappa_{\mu}^{2})^{2}} \Theta(1 -
    \kappa_{\mu})\right]
\end{equation}
using the Heaviside step function $\Theta$, where there is implicitly
no double-counting at exactly $\kappa_{\mu} = 1$. This bound depends
on the singular values $\kappa_{\mu}$ of
$\Im(\hat{V}_{2}^{-1\star})^{-1/2} \hat{Y}^{(0)}_{3}
\Im(\hat{V}_{1}^{-1\star})^{-1/2}$, which combines information about
dissipation in components 1 \& 2 from
$\Im(\hat{V}_{n}^{-1\star})^{-1/2}$ for $n \in \{1, 2\}$ with
information about propagation of forces through component 3 in
$\hat{Y}^{(0)}_{3}$. This bound may be more general if these
contributions, primarily involving the material response for the
former two operators and geometric effects for the latter operator
(which is certainly true of RHT~\cite{MoleskyPRB2020,
  VenkataramPRL2020}), could be separated. Such a separation is made
possible~\cite{Hogben2013} by the inequality $\sigma_{i}
(\hat{M}\hat{N}) \leq \left\lVert \hat{M} \right\rVert_{2} \sigma_{i}
(\hat{N})$ for any operators $\hat{M}$ and $\hat{N}$, where
$\sigma_{i}(\hat{O})$ refers to the $i$th singular value of operator
$\hat{O}$ arranged in a consistent (either nonincreasing or
nondecreasing) order, and $\left\lVert \hat{O} \right\rVert_{2}$ is
the subordinate 2-norm, namely the largest singular value, of the
operator $\hat{O}$; the inequality $\sigma_{i} (\hat{M} \hat{N}) \leq
\sigma_{i} (\hat{M}) \left\lVert \hat{N} \right\rVert_{2}$ follows by
replacing $\hat{M}\hat{N}$ with its Hermitian adjoint, which does not
affect the singular values of that product or its factors. Applying
this repeatedly gives the inequality $\kappa_{\mu}^{2} \leq \zeta_{1}
\zeta_{2} g_{\mu}^{2}$, where $g_{\mu}$ is the corresponding singular
value of the operator $\hat{P}(\hat{V}_{2}) \hat{Y}^{(0)}_{3}
\hat{P}(\hat{V}_{1})$, while $\zeta_{n} = \left\lVert
\Im(\hat{V}_{n}^{-1\star})^{-1} \right\rVert_{2}$ for $n \in \{1,
2\}$. The above bound is monotonically nondecreasing for each
$\kappa_{\mu}$, so plugging in larger values, namely $\sqrt{\zeta_{1}
  \zeta_{2}} g_{\mu}$, in place of $\kappa_{\mu}$ can only loosen the
bound.

\subsection{Generality of singular value bounds}

\begin{figure*}[t!]
  \centering
  \includegraphics[width=0.9\textwidth]{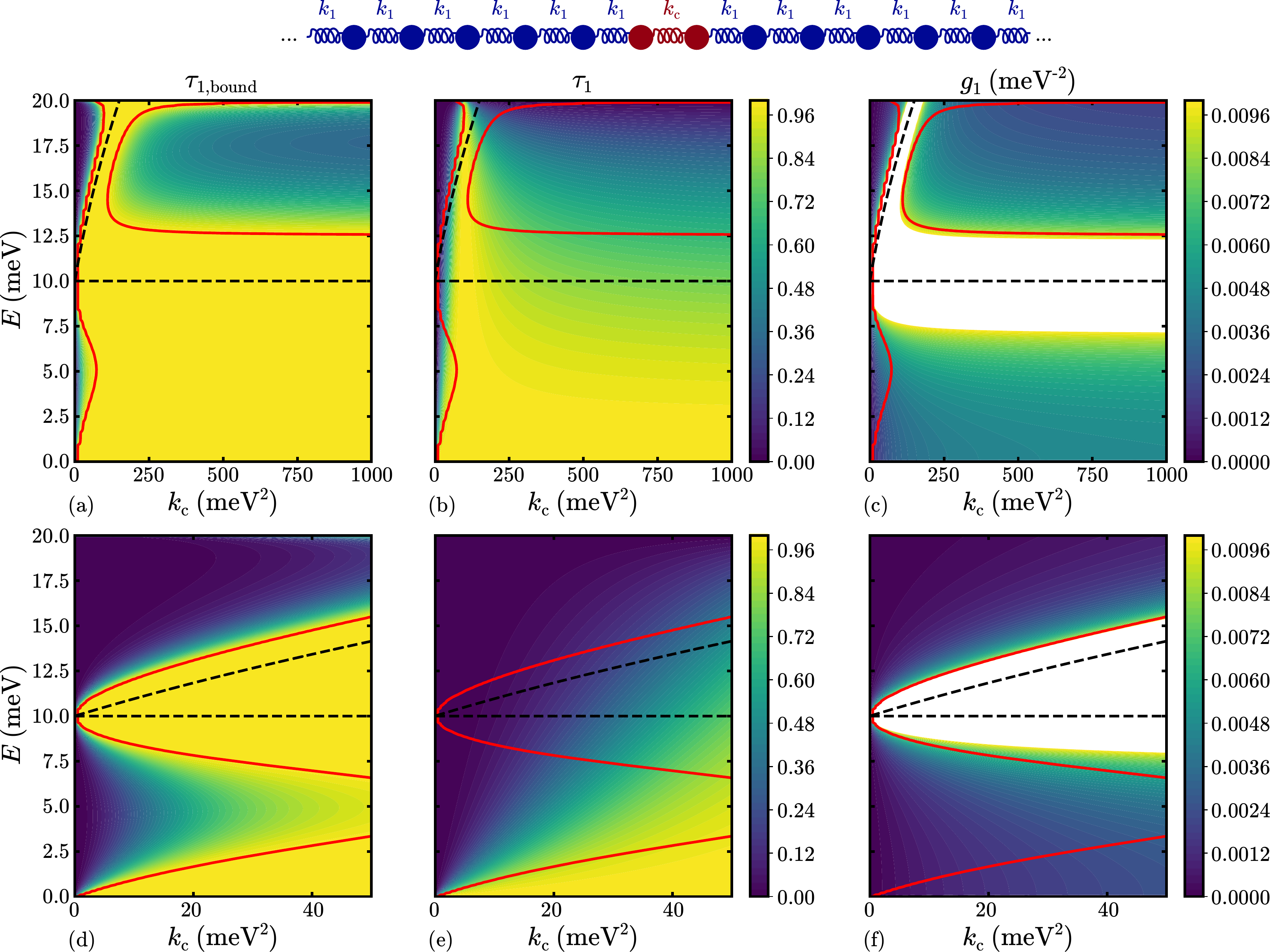}
  \caption{\textbf{1D chain.} Above: system schematic. Below: plots of
    the bound on $\tau_{1}$ (a), as well as the actual values of
    $\tau_{1}$ (b) and $g_{1}$ (c), as functions of $E$ and
    $k_{\mathrm{c}}$. In (a, b), the color bars are shared. (d, e, f)
    Same as (a, b, c) respectively, zoomed in for smaller
    $k_{\mathrm{c}}$. In all plots, the same vertical axis (for $E$)
    is used, and the dashed black lines are $E = \sqrt{k_{1}}$ (lower,
    horizontal) and $E = \sqrt{k_{1} + 2k_{\mathrm{c}}}$ (upper,
    curved), corresponding respectively to the upper and lower
    frequency eigenvalues of the uncoupled junction where $g_{1}$
    diverges and therefore where the bound on $\tau_{1}$ is guaranteed
    to reach unity. Additionally, the red contours denote where
    $\tau_{1,\mathrm{bound}} = 0.99$, as identified from (a) or (d),
    and thereby indicate where $\tau_{1,\mathrm{bound}} < 1$.}
  \label{fig:1Dchainresults}
\end{figure*}

To summarize, while the Landauer bound $\Phi_{\mathrm{L}} = \sum_{\mu}
1$ assumes saturation of the transmissivity for each channel, our
bound shows that this is generally not possible, with $\Phi \leq
\Phi_{\mathrm{opt}} \leq \Phi_{\mathrm{L}}$ at each frequency; our
bound shows not only how many channels may contribute, but also what
the maximum transmissivity for each channel may be that is even
tighter than the prior upper limit of unity. Specifically, we find
that
\begin{equation} \label{eq:Phiopt}
  \Phi_{\mathrm{opt}} = \sum_{\mu} \Bigg[\Theta(\zeta_{1} \zeta_{2}
    g_{\mu}^{2} - 1) + \frac{4\zeta_{1} \zeta_{2} g_{\mu}^{2}}{(1 +
      \zeta_{1} \zeta_{2} g_{\mu}^{2})^{2}} \Theta(1 - \zeta_{1}
    \zeta_{2} g_{\mu}^{2})\Bigg]
\end{equation}
depends on the ``material response factors'' $\zeta_{n} = \left\lVert
\Im(\hat{V}_{n}^{-1\star})^{-1} \right\rVert_{2}$ for $n \in \{1,
2\}$, and the ``transmissive efficacies'' $g_{\mu}$ defined as the
singular values of the operator $\hat{P}(\hat{V}_{2})
\hat{Y}^{(0)}_{3} \hat{P}(\hat{V}_{1})$; there is implicitly no
double-counting at exactly $\zeta_{1} \zeta_{2} g_{\mu}^{2} = 1$, and
a ``recipe'' explaining how to practically compute these bounds for
CHT is provided in~\appref{glossary}. Thus, our bounds capture, per
channel, the interplay between the material response of components 1
and 2 with the transmission properties of component 3 in isolation
between its subparts coupled to each of the other components. The
contribution to each channel $\mu$ is at least as tight as the
per-channel Landauer limit of unity, and only approaches the Landauer
limit if the material response factors, representing a combination of
the inverse dissipation of components 1 or 2 and the coupling of that
component to component 3, are large enough compared to the
transmissive efficacies $g_{\mu}$ for each channel $\mu$. For the
particular case of RHT~\cite{MoleskyPRB2020, VenkataramPRL2020}, as
$\hat{Y}^{(0)}_{3}$ represents the known vacuum Maxwell Green's
function in all of space, broader statements can be made with respect
to domain monotonicity, generality with respect to geometry, and so
on, but for other forms of heat transfer, component 3 may have
specific material properties and shapes that preclude broader
statements along those lines. In any case, these bounds are guaranteed
to be at least as tight as the Landauer bounds, and can in principle
be much tighter, as we demonstrate for the case of phonon CHT in a
representative system in the following section.

\section{Phonon Heat Transfer across a 1D chain}
\label{sec:results}

In this section, we apply our limits to the simple but representative
system depicted in \figref{1Dchainresults}(a). Specifically, we
consider phonon transport in the longitudinal direction in a
1-dimensional (1D) chain, comparing directly to results by
Kl\"{o}ckner et al~\cite{KlocknerPRB2018}, using the same conventions
that $\hbar = m = 1$, and that the central junction is made of two
atoms coupled at strength $k_{\mathrm{c}}$ to each other and at
strength $k_{1}$ to the respective leads (which have uniform internal
couplings $k_{1}$). Our unit convention means that $\omega$ is in
units of $\mathrm{meV}$, while $k_{\mathrm{c}}$ and $k_{1}$ are in
units of $\mathrm{meV}^{2}$; in particular, consistent with that work,
we set $k_{1} = 100~\mathrm{meV}^{2}$ for ease of comparison. The
analysis in that prior work shows that dissipation vanishes for $E
\geq 2\sqrt{k_{1}}$ and $E = 0$, and so we restrict consideration to
$E \in (0, 2\sqrt{k_{1}})$; additionally, straightforward algebraic
manipulations yield the figures of merit, $\zeta_{1} = \zeta_{2} =
\frac{2k_{1}^{2}}{E\sqrt{4k_{1} - E^{2}}}$ and $g_{1} =
\frac{k_{\mathrm{c}}}{|(E^{2} - (k_{\mathrm{c}} + k_{1}))^{2} -
  k_{\mathrm{c}}^{2}|}$.

It can be seen in~\figref{1Dchainresults}(a--c) that for
$k_{\mathrm{c}} = k_{1}$, perfect transmission is possible in
actuality, and the bounds reflect this. Such a rate-matching condition
corresponds to the ``defect'' in the central junction no longer
behaving distinctly from the leads, so the infinite 1D chain is
uniform, and phonons can be perfectly transmitted at any
frequency. For $k_{\mathrm{c}} > k_{1}$, although the transmissive
efficacy $g_{1}$ need not be particularly large for such combinations
of $(E, k_{\mathrm{c}})$, the material response factors $\zeta$ are
large enough for the bound to essentially saturate the Landauer limit
of unity. With respect to the actual heat transfer, in this regime,
the central spring is much stiffer than those of the leads, so
low-frequency excitations $E \leq \sqrt{k_{1}}$ perfectly transmit
across the rigid central spring, while high-frequency excitations $E >
\sqrt{k_{1}}$ largely reflect from the defect, so the actual
transmission nearly saturate the bounds too.

Meanwhile, for $k_{\mathrm{c}} < k_{1}$ as seen
in~\figref{1Dchainresults}(d--f), for which decreasing
$k_{\mathrm{c}}$ may be physically interpreted as increasing the
distance between the two leads (associating the closer atom to each
lead in the junction with that lead), for most combinations of $(E,
k_{\mathrm{c}})$, the actual transmission, despite being quite close
to zero, nearly saturates our bound. This is because for such small
$k_{\mathrm{c}}$, most frequencies will lie far from the resonant
modes of the junction in isolation, so the response of the junction is
quite small. Only for $E$ close to the values $\{\sqrt{k_{1}},
\sqrt{k_{1} + 2k_{\mathrm{c}}}\}$ does our bound come close to the
Landauer limit of unity while the actual transmission does not: this
is because these are the resonant frequencies of the junction in
isolation, whereas the actual transmission depends on the response of
the junction dressed by the two leads and their dissipations, though
the range of frequencies over which this deviation occurs narrows as
$k_{\mathrm{c}}$ decreases further.

From this, it can be concluded that the only points where our bounds
deviate significantly from the actual transmission are near resonances
of the junction in isolation, as that is where the transmissive
efficacy diverges whereas the actual transmission depends on the
response in the presence of the leads. Otherwise, our bounds come much
closer to the actual transmission than the Landauer limits of unity at
most combinations of $(E, k_{\mathrm{c}})$.

\section{Concluding remarks}

We have derived new bounds for heat transfer in arbitrary systems with
linear bosonic response, and showed that for particular molecular
junction geometries of interest to phonon CHT in the linear regime,
these per-channel bounds can not only be much tighter than the
per-channel Landauer limits of unity across many frequencies but can
actually approach the true transmission eigenvalues. As the only
points where our bounds approach the Landauer limits but the actual
transmission eigenvalues do not are those corresponding to resonances
of the junction in isolation (where dressing by the dissipation of the
leads matters more), this suggests that in general, our bounds may be
tight when the density of states is relatively low, and that sum rules
on the density of states could therefore lead to sum rules for heat
transfer integrated over all frequencies, a subject for future
work. Additionally, as a particular junction structure defines the
transmissive efficacies $g_{\mu}$ while the leads with the couplings
to the junction define the material response factors $\zeta$, it
should be possible at each frequency to determine for a given junction
what $\zeta$ allows for saturation of the bounds, and then explore
junction designs to arrive at transmissive efficacies $g_{\mu}$ at
each frequency able to come close to saturating the Landauer limits
of unity (subject to the aforementioned sum rules), though we leave
this to future work too.

\emph{Acknowledgments}.---The authors thank Riccardo Messina, Philippe
Ben-Abdallah, and Le\'{o}n Martin for the helpful comments and
suggestions. This work was supported by the National Science
Foundation under Grants No. DMR-1454836, DMR 1420541, DGE 1148900, the
Cornell Center for Materials Research MRSEC (award no. DMR1719875),
the Defense Advanced Research Projects Agency (DARPA) under agreement
HR00111820046, and the Spanish Ministry of Economy and Competitiveness
(MINECO) (Contract No. FIS2017-84057-P). The views, opinions and/or
findings expressed herein are those of the authors and should not be
interpreted as representing the official views or policies of any
institution.

\appendix

\section{Derivation of alternative bounds} \label{sec:alternatebounds}
In this appendix, we derive alternative bounds to heat transfer that
do not rely on any assumptions about the couplings of component 3 to
components 1 \& 2. As discussed in~\cite{VenkataramARXIV2020}, the
energy transfer spectrum can be written as
\begin{equation}
  \Phi = 4~\left\lVert \Im(\hat{Z}^{(0)\star}_{2})^{1/2} \hat{Y}_{2,2}
  \Delta\hat{Z}_{2,1} \hat{Y}_{1} \Im(\hat{Z}^{(0)\star}_{1})^{1/2}
  \right\rVert_{\mathrm{F}}^{2}
\end{equation}
in terms of the Frobenius norm squared $\left\lVert \hat{O}
\right\rVert_{\mathrm{F}}^{2} = \trace{\hat{O}^{\dagger} \hat{O}}$,
having defined the operators $\Delta\hat{Z}_{mn} \equiv
-\Delta\hat{Z}_{m,3} \hat{Y}^{(0)}_{3} \Delta\hat{Z}_{3,n}$ for $m, n
\in \{1, 2\}$, and in terms of these the operators $\hat{Y}_{1} \equiv
(\hat{Z}^{(0)}_{1} + \Delta\hat{Z}_{1,1})^{-1}$ and $\hat{Y}_{2,2}
\equiv (\hat{Z}^{(0)}_{2} + \Delta\hat{Z}_{2,2} - \Delta\hat{Z}_{2,1}
\hat{Y}_{1} \Delta\hat{Z}_{1,2})^{-1}$; we point out that although the
operators $\Delta\hat{Z}_{3,n}$ (and its transpose) are real-valued
for $n \in \{1, 2\}$, the operators $\Delta\hat{Z}_{mn}$ defined above
for $m, n \in \{1, 2\}$ may in general be complex-valued due to the
dependence on $\hat{Y}^{(0)}_{3}$.

Using the definitions in the main text of the relevant quantities
$\ket{x_{n}}$, $\ket{F_{n}}$, and $\ket{F^{(0)}_{n}}$ for $n \in \{1,
2, 3\}$, as well as the definitions of absorption, scattering, and
extinction in the main text, it can be seen that for a general
external force $\ket{F^{(0)}_{1}}$ on component 1 in the presence of
component 3 but not component 2, the scattered power is
$\Phi_{\mathrm{sca}} = \frac{\omega}{2} \bracket{F^{(0)}_{1},
  (\Im(\hat{Y}_{1}) - \hat{Y}_{1}^{\star} \Im(\hat{Z}^{(0)\star}_{1})
  \hat{Y}_{1}) F^{(0)}_{1}}$. Similarly, for a general external force
$\ket{F^{(0)}_{\mathrm{A}}}$ on the aggregate of components 1 \& 2,
the far-field scattering from component 2 (i.e. into component 3) is
$\Phi_{\mathrm{sca}} = \frac{\omega}{2} \bracket{F^{(0)}_{2},
  (\Im(\hat{Y}_{2,2}) - \hat{Y}_{2,2}^{\star}
  (\Im(\hat{Z}^{(0)\star}_{2}) + \Delta\hat{Z}_{2,1}^{\star}
  \hat{Y}_{1}^{\star} \Im(\hat{Z}^{(0)\star}_{1}) \hat{Y}_{1}
  \Delta\hat{Z}_{1,2})\hat{Y}_{2,2})F^{(0)}_{2}}$. This does not
require any further assumptions about the couplings to component 3
because all of these quantities are cast in terms of response
functions of components 1 \& 2, which are assumed to be disjoint, as
opposed to the response functions of the subsets of DOFs of component
3 coupled to each of the other components (which might not be). Thus,
the relevant operators which must be positive-semidefinite are
$\Im(\hat{Y}_{1}) - \hat{Y}_{1}^{\star} \Im(\hat{Z}^{(0)\star}_{1})
\hat{Y}_{1}$, $\Im(\hat{Y}_{2,2}) - \hat{Y}_{2,2}^{\star}
(\Im(\hat{Z}^{(0)\star}_{2}) + \Delta\hat{Z}_{2,1}^{\star}
\hat{Y}_{1}^{\star} \Im(\hat{Z}^{(0)\star}_{1}) \hat{Y}_{1}
\Delta\hat{Z}_{1,2})\hat{Y}_{2,2}$, and their respective transposes
due to reciprocity.

The remainder of the derivation follows exactly analogously to the
main text, with the replacements $\hat{B}_{1} \to
\Im(\hat{Z}^{(0)\star}_{1})^{1/2} \hat{Y}_{1}
\Im(\hat{Z}^{(0)\star}_{1})^{1/2}$, $\hat{B}_{2} \to
\Im(\hat{Z}^{(0)\star}_{2})^{1/2} \hat{Y}_{2,2}
\Im(\hat{Z}^{(0)\star}_{2})^{1/2}$, and $\hat{A} \to
\Im(\hat{Z}^{(0)\star}_{2})^{1/2} \Delta\hat{Z}_{2,1}
\Im(\hat{Z}^{(0)\star}_{1})^{1/2} \hat{B}_{1} \times
\hat{B}_{1}^{\star} \Im(\hat{Z}^{(0)\star}_{1})^{1/2}
\Delta\hat{Z}_{1,2}^{\star} \Im(\hat{Z}^{(0)\star}_{2})^{1/2}$; these
follow all of the same requisite properties as their counterparts in
the main text. Therefore, the bound can again be written as $\Phi \leq
\Phi_{\mathrm{opt}} \leq \Phi_{\mathrm{L}}$ with $\Phi_{\mathrm{opt}}$
given in~\eqref{Phiopt}, redefining notation regarding the singular
values for these operators such that $\zeta_{n} = \left\lVert
\Im(\hat{Z}^{(0)\star}_{n})^{-1} \right\rVert_{2}$ and $g_{\mu}$ are
now the singular values of $\Delta\hat{Z}_{2,1} = -\Delta\hat{Z}_{2,3}
\hat{Y}^{(0)}_{3} \Delta\hat{Z}_{3,1}$. Once again, this form of our
bound has the benefit of being evaluable even if some DOFs of
component 3 are simultaneously coupled to components 1 \&
2. Additionally, the material response factors $\zeta_{n}$ depend only
on the properties of components 1 \& 2 in isolation, without any
reference to couplings. However, there are two points that may be
practical drawbacks. The first is that the transmissive efficacies
$g_{\mu}$ depend on both the coupling strengths and the properties of
component 3 in isolation, though these effects can be disentangled by
further bounding $g_{\mu} \leq \left\lVert \Delta\hat{Z}_{2,3}
\right\rVert_{2} \left\lVert \Delta\hat{Z}_{3,1} \right\rVert_{2}
\sigma_{\mu} (\hat{P}(\hat{V}_{2}) \hat{Y}^{(0)}_{3}
\hat{P}(\hat{V}_{1}))$ as an extension of the steps in the derivation
in the main text. The second is that particularly in phonon CHT, a
system of broad interest takes components 1 \& 2 to be semi-infinite
leads, with component 3 being a small junction. This means that the
procedure in the main text yields material response factors
$\zeta_{n}$ that can be easily computed from small matrices, as the
matrices $\Delta\hat{Z}_{3,n} \hat{Y}^{(0)}_{n} \Delta\hat{Z}_{n,3}$
can be computed through decimation or similar procedures; by contrast,
the procedure in this appendix requires the full matrices
$\Im(\hat{Z}^{(0)\star}_{n})$ for $n \in \{1, 2\}$, which are large
and might technically vanish unless dissipation is added by hand.

\section{Glossary of relevant quantities for bounds on CHT} \label{sec:glossary}

As the quantities discussed in this manuscript are quite general, it
is useful to draw specific correspondences to operators common to
nonequilibrium Green's function analyses of CHT in order to more
clearly explain how to compute these bounds to CHT in practice. For
reference, the notation we use is generally consistent with notation
for phonon CHT in several prior works~\cite{BurklePRB2015,
  KlocknerPRB2016, KlocknerPRB2017A, KlocknerPRB2017B,
  KlocknerPRB2017C, KlocknerPRB2018}; analogous bounds can be applied
to electron CHT with appropriate replacements of operators. The
following is a set of steps that can be used as a ``recipe'' for
computing the bounds on CHT developed in this manuscript.

\begin{enumerate}
\item Ensure that all quantities have consistent units. For instance,
  for consistency with prior works~\cite{BurklePRB2015,
    KlocknerPRB2016, KlocknerPRB2017A, KlocknerPRB2017B,
    KlocknerPRB2017C, KlocknerPRB2018} on phonon CHT, it will be
  assumed that all spring constants are normalized by the atomic
  masses and by $\hbar$, such that $K_{ai,bj} = \frac{\hbar^{2}}{m_{a}
    m_{b}} \frac{\partial^{2} E}{\partial x_{ai} \partial x_{bj}}$,
  and that the angular frequency $\omega$ will be replaced by the
  energy $E = \hbar\omega$ as the argument of frequency domain
  response quantities. Furthermore, the components 1 \& 2 will be
  referred to as leads $n \in \{\mathrm{L}, \mathrm{R}\}$, while
  component 3 will be referred to as the central junction
  $\mathrm{C}$.

\item At each $E$, compute $\bs{\Pi}^{\mathrm{r}}_{n} =
  \bs{K}_{\mathrm{C}n} \bs{d}^{\mathrm{r}}_{nn} \bs{K}_{n\mathrm{C}}$,
  where $\bs{d}^{\mathrm{r}}_{nn} = ((E + \im\eta)\bs{1}_{nn} -
  \bs{K}_{nn})^{-1}$ is given in terms of an infinitesimal real
  parameter $\eta$ to yield a finite dissipation in each lead $n$.

\item With this, compute $\zeta_{n} = \left\lVert
  \asym(\bs{\Pi}^{\mathrm{r}\dagger -1}_{n})^{-1}
  \right\rVert_{2}$. That is, compute the standard Hermitian adjoint
  of the matrix $\bs{\Pi}^{\mathrm{r}}_{n}$, then take the inverse of
  that within the subspace of DOFs of the junction $\mathrm{C}$ that
  are coupled to the given lead $n$, then compute the anti-Hermitian
  part $\asym(\bs{\Pi}^{\mathrm{r}\dagger -1}_{n})$ (though note that
  the anti-Hermitian part is by definition a Hermitian operator), then
  compute the smallest singular value in that subspace, and set
  $\zeta_{n}$ equal to the reciprocal of that smallest singular
  value. This requires nontrivial dissipation, so
  $\asym(\bs{\Pi}^{\mathrm{r}}_{n})$ should not vanish.

\item At each $E$, compute $\bs{d}^{\mathrm{r}}_{\mathrm{CC}} = ((E +
  \im\eta)\bs{1}_{\mathrm{CC}} - \bs{K}_{\mathrm{CC}})^{-1}$; note
  that this is the response of the \emph{uncoupled} junction, which is
  \emph{not} the same as $\bs{D}^{\mathrm{r}}_{\mathrm{CC}}$.

\item Construct the off-diagonal block $\bs{P}_{\mathrm{C(R)}}
  \bs{d}^{\mathrm{r}}_{\mathrm{CC}} \bs{P}_{\mathrm{C(L)}}$. That is,
  extract the off-diagonal block of
  $\bs{d}^{\mathrm{r}}_{\mathrm{CC}}$ where the rows correspond to
  atoms in the central junction $\mathrm{C}$ with nonzero couplings to
  the right lead $\mathrm{R}$, and the columns correspond to atoms in
  the central junction $\mathrm{C}$ with nonzero couplings to the left
  lead $\mathrm{L}$. Note that this assumes that no atoms in the
  central junction couple simultaneously to both leads, so single-atom
  junctions cannot be treated as single atoms per se (i.e. the
  junction needs to be artificially increased in size to include more
  atoms in the leads until those overlaps disappear).

\item Find the singular values $g_{\mu}$ of this off-diagonal block
  $\bs{P}_{\mathrm{C(R)}} \bs{d}^{\mathrm{r}}_{\mathrm{CC}}
  \bs{P}_{\mathrm{C(L)}}$; the label $\mu$ is said to denote the
  channel. Note that this off-diagonal block is generally not square
  (i.e. it might not be the case that the numbers of atoms in the
  junction coupling to each of the leads are the same), but the
  singular value decomposition (SVD) will always exist and should
  always yield real nonnegative values (barring unexpected numerical
  problems).

\item At this $E$, plug the quantities $\zeta_{n}$ and $g_{\mu}$
  into~\eqref{Phiopt} for each channel $\mu$ to yield the bound
  $\Phi_{\mathrm{opt}}$; note the change in labels $n \in \{1, 2\} \to
  \{\mathrm{L}, \mathrm{R}\}$.
\end{enumerate}

\nocite{apsrev41Control} \bibliographystyle{apsrev4-1}

\bibliography{boundspaper}
\end{document}